\documentclass[12pt]{article}
\usepackage{color}
\usepackage{latexsym}
\usepackage{epsfig,amssymb,euscript, mathrsfs}
\usepackage{amsmath}
\definecolor{MyDarkBlue}{rgb}{0.15,0.15,0.45}
\usepackage[linktocpage=true]{hyperref}
\hypersetup{
colorlinks=true,
citecolor=MyDarkBlue,
linkcolor=MyDarkBlue,
urlcolor=MyDarkBlue,
pdfauthor={Luis F Alday, Paul Richmond, James Sparks},
pdftitle={The holographic supersymmetric Renyi entropy in five dimensions},
pdfsubject={hep-th}
}
\newsavebox{\ns}
\newsavebox{\dbrane}
\newsavebox{\dbshort}

\def\be{\begin{equation}}
\def\ee{\end{equation}}
\def\bea{\begin{eqnarray}}
\def\eea{\end{eqnarray}}

\newcommand{\nn}{\nonumber}

\newcommand\R{\mathbb{R}}
\newcommand\Z{\mathbb{Z}}

\newcommand\C{\mathbb{C}}

\newcommand\diff{\mathrm{d}}

\newcommand{\dd}{\mathrm{d}}

\newcommand{\ii}{\mathrm{i}}

\newcommand{\ex}{\mathrm{e}}
\newcommand{\vol}{\mathrm{vol}}

\newlength{\sswidth}

\newcommand{\bbeta}{\gamma}
\newcommand{\taunew}{\tau}
\newcommand{\ang}{\alpha}

\numberwithin{equation}{section}       

\begin{document}

\bibliographystyle{utphys}

\begin{titlepage}

\begin{center}

\today

\vskip 2.3 cm 

\vskip 5mm

{\Large \bf The holographic supersymmetric R\'enyi entropy }
\vskip 0.5cm
{\Large \bf  in five dimensions }

\vskip 15mm

{Luis F. Alday, Paul Richmond and James Sparks}

\vspace{1cm}
\centerline{{\it Mathematical Institute, University of Oxford,}}
\centerline{{\it Andrew Wiles Building, Radcliffe Observatory Quarter,}}
\centerline{{\it Woodstock Road, Oxford, OX2 6GG, UK}}

\end{center}

\vskip 2 cm

\begin{abstract}
\noindent  We compute the supersymmetric R\'enyi entropy across an entangling three-sphere
for five-dimensional superconformal field theories using localization. 
For a class of $USp(2N)$ gauge theories we construct 
a holographic dual 1/2 BPS black hole solution of  Euclidean Romans $F(4)$ supergravity. 
The large $N$ limit of the gauge theory results agree perfectly with the 
supergravity computations. 

\end{abstract}

\end{titlepage}

\pagestyle{plain}
\setcounter{page}{1}
\newcounter{bean}
\baselineskip18pt
\tableofcontents




\section{Supersymmetric R\'enyi entropy}\label{sec:fieldtheory}

\subsection{R\'enyi entropy in CFT}

Given a quantum field theory an interesting observable is the R\'enyi entropy. 
To define this one divides a spatial slice $\Sigma$ into a region $A$ and its complement 
$B=\Sigma\setminus A$. The Hilbert space then factorizes
\bea
\mathcal{H}	&\cong & \mathcal{H}_A\otimes \mathcal{H}_{B}~.
\eea
The  reduced density matrix $\rho_A$ is  defined as
\bea
\rho_A & = & \mathrm{Tr}_{B} |\, 0\, \rangle\langle \, 0\, |~,
\eea
where $|\, 0\, \rangle$ is the ground state of the theory. 
For any positive integer $n>1$, the R\'enyi entropy $S_n(A)$ associated to $A$ is then defined as
\bea
S_n(A) &=& \frac{1}{1-n}\log\, \frac{\mathrm{Tr}_A \rho^n_A}{\left(\mathrm{Tr}_A \rho_A\right)^n}~.
\eea
This is a refinement of the entanglement entropy, which arises by analytically continuing in 
$n$ and taking the limit
\bea
S_{\mathrm{EE}}(A) &=& \lim_{n\rightarrow 1} S_n(A)~.
\eea

One can define the R\'enyi entropy using the path integral 
formalism as follows.
Consider a Euclidean spacetime with coordinates $(t_E,x,\vec{z})$, where $t_E=\ii t$ is the Euclidean time
and the spatial slice $\Sigma=\{t_E=0\}$. 
The coordinate $x$ is then defined such that
 $x \geq 0$ is the region $A$, and $x<0$ its complement $B$. 
The ground state wave function is given by the path integral
\bea
\Psi[\psi_0(x,\vec{z})] &=& \int_{\displaystyle\left.\psi\right|_{t_E=0}=\psi_0} {\cal D}\psi \, \ex^{-I_E(\psi)} \, ,
\eea
where the fields $\psi$ we are integrating over are defined for negative Euclidean time (or positive imaginary Minkowskian time), 
and $I_E$ is the Euclidean action. The factorization of the $t_E=0$ slice into 
$A\cup B$ leads to  a factorization of the  boundary data
\bea
\psi_0(x,\vec{z})&=& \begin{cases}\ \psi_A(x,\vec{z}) &  ~~~\textrm{for}~ x\geq 0 \, ,\\ 
\ \psi_B(x,\vec{z}) &  ~~~\textrm{for}~ x<0 \, . \end{cases}
\eea
The reduced density matrix is then
\bea
\rho_A(\psi_A^{+},\psi_A^{-}) &=& \int {\cal D}\psi_B \, \Psi^\dagger[\psi_A^{+},\psi_B]\Psi[\psi_A^{-},\psi_B] \, .
\eea
If we let the imaginary time in the two path integral definitions of $\Psi$ run from $0$ to $\pm \infty$ respectively, the density matrix becomes the path integral over fields defined on the full Euclidean space, with the $(t_E,x)$ plane cut along the $x>0$ ray and with $\psi_A$ taking values $\psi_A^{\pm}$ above and below the cut, respectively. The trace of the density matrix is obtained by equating the fields across the cut and carrying out the unrestricted Euclidean path integral. More generally this construction shows that
\bea
\mathrm{Tr}_A \rho^n_A &=& Z_n \, ,
\eea
where $Z_n$ is given by the Euclidean path integral over an $n$-sheeted covering of the cut spacetime. This formulation of the R\'enyi entropy is known as the replica trick \cite{Callan:1994py}, and 
leads to the formula
\bea\label{SnZn}
S_n(A) &=& \frac{1}{1-n}\log\, \frac{Z_n}{(Z_1)^n}~.
\eea

The calculability of $S_n(A)$ depends on the choice of spacetime and region $A$. A natural choice is a spacetime of the form $\R_{t_E} \times \R^{d-1} = \R^d$ and $A$ the unit ball inside $\R^{d-1}$, so that $\partial A = S^{d-2}$. The  metric is
\bea
\dd s^2_{\R^{d}} &=&\dd t_E^2+\dd\rho^2+\rho^2 \dd s^2_{S^{d-2}} \, ,
\eea
where $\dd s^2_{S^{d-2}}$ denotes the round metric on the unit $(d-2)$-sphere.
The region $A$ is the ball $0\leq\rho\leq 1$. For conformal theories it is convenient 
\cite{Calabrese:2004eu}
to perform the computation in the conformally equivalent space $S^d$ with metric
\bea\label{Sd}
\diff s^2_{S^d} &=& \cos^2 \ang\, \diff \taunew^2 + \diff\ang^2 + \sin^2\ang \, \diff s^2_{S^{d-2}}~,
\eea
where  the change of coordinates is
\bea
t_E &=& \frac{\cos\ang \sin \tau}{1+\cos \ang \cos \tau} \, , \\
\rho &=& \frac{\sin \ang}{1+\cos \ang \cos \tau} \, .
\eea
Here $0 \leq \ang \leq \pi/2$ and $\tau$ is periodic with period $2\pi$. In these coordinates the branch locus is at $\ang=\pi/2$ and the cut is at $\tau=0$. In order to compute the R\'enyi entropy we need to evaluate the partition function on the $n$-branched $d$-sphere, in which case the periodicity of $\tau$ is $2\pi n$.

The above replica trick, in which one studies field theory on a singular space, is a convenient method to compute entanglement entropies in conformal field theories. However, if one is interested in constructing holographic duals this singularity persists into the bulk, where gravity becomes dynamical. This raises the issue of how to treat the singularity in gravity 
\cite{Headrick:2010zt}. An ingenious way to circumvent this problem is to instead conformally map the space to $S^1 \times \mathbb{H}^{d-1}$ \cite{Casini:2011kv}
\bea\label{S1H}
\dd s^2_{S^1 \times \mathbb{H}^{d-1}} &=& \dd\tau^2+\frac{\dd q^2}{1+q^2} +q^2  \dd s^2_{S^{d-2}} \, ,
\eea
where $q=\tan \ang$ takes the range $q\in[0,\infty)$. The coordinates in (\ref{S1H}) realize the hyperbolic space $\mathbb{H}^{d-1}$ in a spherical slicing. The branch cut at $\ang=\pi/2$ has now moved to $q=\infty$. In \cite{Casini:2011kv} it was argued that the entanglement entropy maps to a thermal entropy in this space, where the new Euclidean time $\tau$ has period $\beta=2\pi n$, the inverse temperature. The holographic duals are then naturally black hole solutions with hyperbolic horizons (so called topological black holes).

\subsection{Supersymmetry and localization}\label{sec:SUSY}

In \cite{Nishioka:2013haa} the authors studied a supersymmetric version 
of the above R\'enyi entropy for $\mathcal{N}=2$ supersymmetric theories 
on the round three-sphere with $d=3$. This is similarly obtained by 
computing the partition function on $S^3$ branched $n$ times over 
the $S^1$ at $\ang=\pi/2$, but in addition one needs to turn on an appropriate
background R-symmetry gauge field to preserve supersymmetry. After a lengthy computation using localization 
they find that the 
partition function $Z_n$  is simply the partition function of 
the \emph{squashed} sphere $S^3_b$, with squashing parameter $\sqrt{\frac{b_1}{b_2}}=\sqrt{n}$. 
In this section we give a simple explanation for this result, 
which works in general dimensions. Since 
we will be interested mainly  in dimension $d=5$, 
we shall present the argument for this case.

We write the metric on $S^5$ in the form (\ref{Sd}), where we choose coordinates 
on $S^{d-2}=S^3$ as
\bea\label{S3coords}
\dd s^2_{S^3} &=& \dd \theta^2 +\cos^2\theta\dd\psi^2 + \sin^2\theta\dd\phi^2~.
\eea 
Here $\psi$ and $\phi$ both have period $2\pi$, while $0\leq \theta\leq \pi/2$. 
In order to define supersymmetric field theories on $S^5$ (or its branching 
along $S^3$) one needs to choose a Killing spinor $\epsilon$. 
The Killing spinors on $S^5$ have charges 
$\pm{1}/{2}$ under the Lie derivatives along
$\partial_\tau$, $\partial_{\psi}$, $\partial_\phi$, which 
generate a $U(1)^3\subset SO(6)$ subset of isometries. In particular 
our choice of spinor will be such that
\bea\label{charge12}
\mathcal{L}_{\partial_\tau}\epsilon &=& -\frac{\ii}{2}\epsilon~.
\eea
This charge  guarantees that the spinor 
$\epsilon$ is smooth at $\alpha=\pi/2$, where 
$\partial_\tau=0$. Indeed, the normal 
space to $\alpha=\pi/2$ is  a copy of $\R^2\cong \C$. 
One can then introduce a polar radial variable $R=\pi/2-\alpha$, 
and corresponding Cartesian coordinates $X=R\cos\tau$, 
$Y=R\sin\tau$ on this normal space. The 
frame $e_1=\dd X$, $e_2=\diff Y$ rotates with charge 1 
under $\partial_\tau$, so that positive and negative chirality 
spinors in $\R^2$ correspondingly rotate with charge $\pm {1}/{2}$, 
respectively. One could instead move to the non-rotating 
frame $\hat{e}_1=\diff R$, $\hat{e}_2=R\diff\tau$, 
in which the spinor will then have an explicit overall phase
$\ex^{-\ii\tau/2}$. However, this frame is singular at the origin $R=0$
which is  why the spinor looks singular there.

Similarly, we choose conventions so that $\epsilon$ 
has charge $-{1}/{2}$ under $\partial_\psi$ 
and $+{1}/{2}$ under $\partial_\phi$. The vector bilinear 
$K^\mu=\epsilon^\dagger\gamma^\mu\epsilon$ 
is then the Killing vector
\bea\label{Ksphere}
K &=& -\partial_\phi+\partial_\psi+\partial_\tau~,
\eea
which generates a Hopf foliation of $S^5$.

To form the $n$-branched sphere one simply takes $\tau$ to have period $2\pi n$. 
We may introduce a local complex coordinate $w=R\ex^{\ii\tau}$ on 
the normal space $\R^2$ to the branch locus $\alpha=\pi/2$. 
Then $z=w^{1/n}=R^{1/n}\ex^{\ii\varphi_3}$ 
has $\arg z=\varphi_3=\tau/n$, which has the canonical period $2\pi$.
Moreover, a function is \emph{smooth} at the branch point $R=0$
means that it is smooth in the coordinate $z$. For example, 
when we come to discuss the computation of one-loop determinants 
below, it is convenient to expand in Fourier modes of  the 
$(S^1)^3=U(1)^3$ isometry, and a complete set of modes in the $\tau$ direction 
is then $\ex^{\ii m_3\varphi_3}$, with $m_3\in\Z$. 

The Killing spinor $\epsilon$ on the round 
sphere  has charge $-{\ii n}/{2}$ under $\partial_{\varphi_3}$, 
and is thus singular along the branch locus when $n>1$. 
We may remedy this, as in \cite{Nishioka:2013haa}, 
by introducing the background R-symmetry gauge field
\bea\label{AR}
\mathcal{A} &=& -\frac{n-1}{n}\dd\tau \ =\ -(n-1)\diff\varphi_3~.
\eea
In a five-dimensional supersymmetric gauge theory we view this 
as embedded in $U(1)_R\subset SU(2)_R$, where the gauge covariant derivative 
on $\epsilon$ is\footnote{The second spinor in the $SU(2)_R$ doublet then has the opposite charge under $\mathcal{A}_\mu$.}
\bea\label{gaugecov}
D_\mu \epsilon &=& \nabla_\mu \epsilon + \frac{\ii}{2}\mathcal{A}_\mu\epsilon~.
\eea
The flat gauge field (\ref{AR}) is smooth everywhere on the branched sphere, except at the 
branch locus $\alpha=\pi/2$ where $\partial_\tau=0$. This singularity 
is designed precisely so as to render the resulting Killing spinor smooth. To see this,
note that we may write $\mathcal{A}=\ii g^{-1}(\diff g)=\ii \diff \log g$, where 
\bea
g &=& \ex^{\ii (n-1)\tau /n}~.
\eea
The factor of $1/2$ in (\ref{gaugecov}) is chosen to match our Romans supergravity conventions in the next section, 
but in particular this implies that the spinor transforms as 
\bea
\epsilon_{\mathrm{branched}} & = & g^{1/2} \epsilon~.
\eea
Of course then
\bea
D_\mu \epsilon_{\mathrm{branched}} &=& g^{1/2}\nabla_\mu \epsilon~,
\eea
so that $\epsilon_{\mathrm{branched}}$ satisfies the same Killing spinor 
as on the round sphere, but with the Levi-Civita spin
 connection replaced by gauge-covariant derivative (\ref{gaugecov}). 
Moreover, in a non-rotating frame for $\partial_\tau$, the $\tau$-dependent phase 
of the new spinor is
\bea
\ex^{-\ii\tau/2}\cdot \ex^{\ii (n-1)\tau/2n} \ = \ \ex^{-\ii \tau/2n} \ =\ \ex^{-\ii\phi_3/2}~.
\eea
This shows that the charged spinor $\epsilon_{\mathrm{branched}}$ 
is non-singular at the branch locus, and hence non-singular everywhere on the branched five-sphere. 
Moreover, since the Killing vector bilinear $K^\mu=\epsilon_{\mathrm{branched}}^\dagger\gamma^\mu\epsilon_{\mathrm{branched}}$
is the same as that for the uncharged spinor $\epsilon$, we have from (\ref{Ksphere})
\bea
K &=&  -\partial_\phi+\partial_\psi+\partial_\tau \ =\ b_1\partial_{\varphi_1}+b_2\partial_{\varphi_2}+b_3\partial_{\varphi_3}~.
\eea
Here we have introduced the angular coordinates
\bea\label{Fernandoangles}
\varphi_1 &=& -\phi\, ,\qquad \varphi_2 \ = \ \psi \, , \qquad \varphi_3 \ = \ \frac{1}{n}\tau\, ,
\eea 
on $U(1)^3$, which all have canonical $2\pi$ periods, and $(b_1,b_2,b_3)=(1,1,\frac{1}{n})$. 

Imagine now computing the perturbative partition function of a supersymmetric field theory 
on the $n$-branched five-sphere using localization. Locally this computation is the same as that on the round sphere \cite{Kallen:2012va, Kim:2012ava}. What changes are the boundary conditions along the 
branch locus at $\alpha=\pi/2$. However, as explained above, these boundary conditions 
simply mean that fields are smooth in the $z$ coordinate, rather than the original $w$ 
coordinate. In particular,  one expands in Fourier modes $\exp[\ii(m_1\varphi_1+m_2\varphi_2+m_3\varphi_3)]$ where $m_i\in\Z$, $i=1,2,3$. 
The charge of such a mode under the supersymmetric Killing vector $K$ is 
then $m_1b_1+m_2b_2+m_3b_3$. The Killing spinor $\epsilon_{\mathrm{branched}}$ itself similarly 
has charge $-(b_1+b_2+b_3)/2$. Combining these observations with the 
structure of the one-loop calculations in three dimensions in \cite{Hama:2011ea, Alday:2013lba} 
then leads immediately to the result in \cite{Nishioka:2013haa}:\ the partition function $Z_n$ is simply the partition function 
of any three-sphere background with $(b_1,b_2)=(1,\frac{1}{n})$. 
In particular most modes in the one-loop determinant 
pair under supersymmetry, so that their contributions cancel. 
For the remaining unpaired modes, their eigenvalues depend on the 
background geometry only via their charge $m_1b_1+m_2b_2$. 
The determinant over normalizable modes then leads to a double 
sine function $S_2(\cdot \mid (b_1,b_2))$.
In five dimensions, similar reasoning applies to 
the
 explicit computation 
of the perturbative partition function on $S^5$.  
We then expect the result
\bea
{Z}^{\text{pert}}_n \, = \,  C(\mathbf{b})\prod_{a =1}^{\mathrm{rank} \ G}  \int_{-\infty}^{\infty} \mathrm{d} \sigma_a \, \ex^{ - \frac{(2 \pi)^{3}}{b_1 b_2 b_3} \mathscr{F}(\sigma)}\frac{ \prod_\alpha S_3 \left( - \ii \alpha (\sigma) \mid \mathbf{b} \right)}{\prod_\rho S_3 \left( - \ii \rho (\sigma) + \tfrac{1}{2}(b_1+b_2+b_3)
\mid \mathbf{b}  \right)} \, . \label{Z}
\eea
Here the prefactor $C(\mathbf{b})$  depends only on $\mathbf{b}=(b_1,b_2,b_3)=(1,1,\frac{1}{n})$, and 
in particular will not contribute to the large $N$ limit of interest in the next subsection. 
The product over $\alpha$ in the numerator 
is over roots of the gauge group $G$, while the product over $\rho$ in the denominator is over weights 
in a weight space decomposition of the matter representation $\mathbf{R}$. 
The integral in $\sigma_a$ is over the Cartan of $G$, $\mathscr{F}$ is the prepotential 
of the theory, while 
$S_3 ( \cdot \mid \mathbf{b} )$ is  the triple sine function. 
The result (\ref{Z}) also agrees
with the conjecture the authors made in \cite{Alday:2014rxa,  Alday:2014bta}:\ that the partition function for any supersymmetric five-sphere background 
depends on the background only via the Killing vector $K$. 
In particular, (\ref{Z}) equals the squashed five-sphere perturbative partition 
function \cite{Imamura:2012bm}.

\subsection{Large $N$ limit of $USp(2N)$ superconformal theories}\label{sec:largeN}

The result for the perturbative partition function (\ref{Z}) 
is valid for a general supersymmetric gauge theory in five dimensions.
We  now focus on a 
particular class of theories with gauge group 
$G=USp(2N)$ and matter consisting of $N_f$ hypermultiplets in the fundamental and a single hypermultiplet in the anti-symmetric representation of $G$. These theories arise from a system of $N$ D4-branes and some number of D8-branes and orientifold 
planes in massive type IIA string theory, and have a large $N$ limit that has a dual description in massive type IIA supergravity
\cite{Ferrara:1998gv, Brandhuber:1999np, Bergman:2012kr}. 
For these theories, the large $N$ limit of  (\ref{Z}) gives the free energy \cite{Alday:2014bta}
\bea
{F} \ = \ -\log Z_n^{\mathrm{pert}} & =&    \frac{ (b_1+b_2+b_3)^3}{27 b_1 b_2 b_3} {F}_{{S}^{5}_{\mathrm{round}}} \nn\\
&=& \frac{(1+2n)^3}{27n^2}{F}_{{S}^{5}_{\mathrm{round}}}~.
 \label{freeenergyfinal}
\eea
Here ${F}_{{S}^{5}_{\mathrm{round}} } = \frac{9 \sqrt{2} \pi N^{5/2}}{5 \, \sqrt{8-N_f}} + \mathcal{O}\left( N^{3/2} \right) $ is the large $N$ limit of the free energy on the round five-sphere computed in reference \cite{Jafferis:2012iv}. This results
in the following  large $N$ R\'enyi entropy
\bea
S_n \ = \ S_n(S^3) &=& -\frac{1+7n+19n^2}{27n^2}F_{{S}^{5}_\mathrm{round}}~.
\eea
In the next section we will reproduce this result
from the holographic dual computation. 


\section{Holographic dual}\label{sec:gravitydual}

Following \cite{Casini:2011kv}, and similar computations in lower dimensions \cite{Huang:2014gca,Nishioka:2014mwa,Huang:2014pda,Crossley:2014oea}, the holographic supersymmetric 
R\'enyi entropy is computed from a 1/2 BPS Euclidean black hole solution. 
As explained in \cite{Alday:2014bta}, we may construct this dual
solution in Euclidean Romans $F(4)$ supergravity, and then uplift 
this to a solution of massive IIA string theory. 

\subsection{Euclidean Romans $F(4)$ supergravity}

The bosonic fields of the six-dimensional Romans supergravity theory \cite{Romans:1985tw} consist of the metric, a scalar field $X$, a two-form potential $B$, and a one-form potential $A$, together with an $SO(3)\sim SU(2)$ gauge field $A^i$ where $i=1,2,3$. For the solution of interest in this paper the two-form potential vanishes, $B=0$, and we work in a gauge in which the Stueckelberg one-form $A$ is zero. Setting also the gauge coupling constant to unity, the Euclidean equations of motion are \cite{Alday:2014rxa,Alday:2014bta} 
\begin{eqnarray}
\label{FullEOM}
F^i \wedge F^i & =& 0\, , \nn\\
D(X^{-2}*F^i) & = & 0\, ,\nn\\
\diff\left(X^{-1}*\dd X\right) &=& -  \left(\tfrac{1}{6}X^{-6}-\tfrac{2}{3}X^{-2}+\tfrac{1}{2}X^2\right)*1 -\tfrac{1}{8}X^{-2}\left(F^i\wedge *F^i\right) \, .
\end{eqnarray}
The first equation is a remnant of the $B$-field equation of motion, and $D\omega^i=\dd\omega^i -\epsilon_{ijk}A^j\wedge \omega^k$ is the 
$SO(3)$ covariant derivative. 
The Einstein equation is
\begin{eqnarray}
R_{\mu\nu} &=& 4X^{-2}\partial_\mu X\partial_\nu X + \left(\tfrac{1}{18}X^{-6}-\tfrac{2}{3}X^{-2}-\tfrac{1}{2}X^2\right) g_{\mu\nu} \nn \\
 & &+  \tfrac{1}{2}X^{-2}\left((F^i)^2_{\mu\nu}-\tfrac{1}{8}(F^i)^2g_{\mu\nu}\right)\, ,
\end{eqnarray}
where $(F^i)^2_{\mu\nu} = F^i_{\mu\rho} F^i_\nu{}^\rho$. 
The Euclidean action  is
\bea\label{Fullaction}
I_E \ =\ &  -\displaystyle\frac{1}{16\pi G_6}\int_{M_6} &  R*1-4X^{-2} ( \diff X\wedge *\diff X + \tfrac{1}{8} F^i \wedge * F^i )\nn\\
&& -\left(\tfrac{2}{9}X^{-6}-\tfrac{8}{3}X^{-2}-2X^2\right)*1 \, .
\eea
A solution to the above equations of motion is supersymmetric provided there 
exist non-trivial Dirac spinors $\epsilon_I$, $I=1,2$, satisfying 
the following Killing spinor  and dilatino equation
\begin{align}
D_\mu \epsilon_I \ \ =& \ \ \tfrac{\ii}{4\sqrt{2}} ( X + \tfrac{1}{3} X^{-3} ) \Gamma_\mu \Gamma_7 \epsilon_I + \tfrac{1}{16\sqrt{2}}X^{-1} F_{\nu\rho}^i ( \Gamma_\mu{}^{\nu\rho} - 6 \delta_\mu{}^\nu \Gamma^\rho ) \Gamma_7 ( \sigma^i )_I{}^J \epsilon_J \, ,\label{KSE}\\[10pt]
0 \ \ = &\  \ - \ii X^{-1} \partial_\mu X \Gamma^\mu \epsilon_ I + \tfrac{1}{2\sqrt{2}} \left( X - X^{-3} \right) \Gamma_7 \epsilon_I - \tfrac{\ii}{8\sqrt{2}} X^{-1} F^i_{\mu\nu} \Gamma^{\mu\nu} \Gamma_7 ( \sigma^i )_I{}^J \epsilon_J \, .\label{dilatino}
\end{align}
Here  $\Gamma_\mu$ generate the Clifford 
algebra $\mathrm{Cliff}(6,0)$ in an orthonormal frame, and we have defined the chirality operator
 $\Gamma_7 = \ii \Gamma_{012345}$, which satisfies $(\Gamma_7)^2=1$. 
The covariant derivative acting on the spinor is $D_\mu\epsilon_I=\nabla_\mu\epsilon_I+\frac{\ii}{2} A_\mu^i(\sigma^i)_I{}^J\epsilon_J$.

\subsection{$1/2$ BPS black hole solution}

Our starting point is the charged AdS black hole solution of \cite{Cvetic:1999un}. After a Wick rotation and a relabelling of parameters, the solution is
\bea\label{BH}
\dd s^2 & = & \frac{H(r)^{1/2}}{f(r)} \dd r^2 + \frac{9f(r)}{2H(r)^{3/2}} \dd \taunew^2 + r^2 H(r)^{1/2} \dd s^2_{ \mathbb{H}^4} \, ,
\eea
where 
\bea
H(r) &=& 1 + \frac{Q}{r^3} \nn \, , \\
f(r) &=& -1 -\frac{\bbeta}{r^3}+ \frac{2}{9} r^2 H(r)^2 \, .
\eea
The solution depends on the two parameters $Q$ and $\gamma$,  
and $\dd s^2_{ \mathbb{H}^4}$ is the metric of a unit radius hyperbolic space.  As in 
section \ref{sec:fieldtheory} we choose coordinates so that
\bea
\dd s^2_{ \mathbb{H}^4} &=& \frac{1}{(1+q^2)} \dd q^2 + q^2 \big( \dd \theta^2 + \cos^2 \theta \dd \psi^2 + \sin^2 \theta \dd \phi^2 \big) \, .
\eea
{\it Cf.} equations (\ref{S1H}), (\ref{S3coords}).
The remaining fields are
\bea\label{XA}
X(r) &=& H(r) ^{-1/4} \nn \, , \\
\mathcal{A} &\equiv& A^3 \ = \ 3 \sqrt{1- \frac{\bbeta}{Q}} \, \frac{H(r)-1}{H(r)} \dd \taunew +\mu\dd\taunew\, .
\eea
Notice that the parameter $Q$ is necessarily non-zero if $\bbeta\neq0$. We have also 
added a pure gauge term $\mu\dd\taunew$ to $\mathcal{A}$, which as we shall see is required 
in order that the gauge field is non-singular at the horizon.

The metric (\ref{BH}) is asymptotically locally AdS for large $r$. Specifically 
\bea
\dd s^2 &\simeq & \frac{9 \dd r^2}{2r^2} + r^2\left(\dd\taunew^2 +\diff s^2_{\mathbb{H}^4}\right) \, ,
\eea
to leading order as $r\rightarrow\infty$. Moreover, the scalar field $X\rightarrow 1$ while 
$\mathcal{A}\rightarrow \mu\dd\taunew$. Since $\taunew$ will be periodically identified 
in the next subsection, the conformal boundary geometry is 
$S^1\times \mathbb{H}^4$.

The solution is supported by a single component of the $SU(2)$ gauge field, and without loss of generality we have chosen this to lie along the $i=3$ direction. For this choice of gauge the Killing spinor equations for $\epsilon_1$ and $\epsilon_2$ decouple. Moreover, if the fields are all real then the Killing spinor equation for $\epsilon_2$ is simply the charge conjugate of that for $\epsilon_1$ 
\cite{Alday:2014bta}. Hence we can consider only the spinor $\epsilon = \epsilon_1$ which satisfies
\bea
D_\mu \epsilon & =&  \tfrac{\ii}{4\sqrt{2}} ( X + \tfrac{1}{3} X^{-3} ) \Gamma_\mu \Gamma_7 \epsilon + \tfrac{1}{16\sqrt{2}}X^{-1} \mathcal{F}_{\nu\rho} ( \Gamma_\mu{}^{\nu\rho} - 6 \delta_\mu{}^\nu \Gamma^\rho ) \Gamma_7 \epsilon \label{KSEdecoupled} \, , \\[10pt]
0 & = & - \ii X^{-1} \partial_\mu X \Gamma^\mu \epsilon + \tfrac{1}{2\sqrt{2}} \left( X - X^{-3} \right) \Gamma_7 \epsilon - \tfrac{\ii}{8\sqrt{2}} X^{-1} \mathcal{F}_{\mu\nu} \Gamma^{\mu\nu} \Gamma_7 \epsilon \label{Dilatinodecoupled} \, ,
\eea
with $\mathcal{F} = \dd \mathcal{A}$.

The above black hole solution is 1/2 BPS for $\bbeta=0$. To see this we 
introduce the frame
\begin{eqnarray}
e^{0} & =& \frac{H(r)^{1/4}}{f(r)^{1/2}} \dd r~,~ \quad  \ \,  e^{1}\ =\  \frac{3}{\sqrt{2}} \frac{ f(r)^{1/2} }{H(r)^{1/4} }  \dd \taunew ~,\qquad  ~ \,e^{2} \ = \ \frac{r H(r)^{1/4} }{(1+q^2)^{1/2}} \dd q~,\\[8pt]
e^{3} & =& q r H(r)^{1/4} \dd \theta ~,~~~e^{4} \ = \ q r H(r)^{1/4} \cos \theta \dd \psi~,~~~e^{5} \ = \ q r H(r)^{1/4} \sin \theta \dd \phi~,\nn
\end{eqnarray}
and the following basis of six-dimensional gamma matrices
\bea
\Gamma_0 & = &  \left( \begin{array}{cc}  0 & 1_4 \\ 1_4 & 0 \end{array} \right) \, , \quad \Gamma_m \  = \  \left( \begin{array}{cc}  0 & \ii  \gamma_m \\ - \ii \gamma_m & 0 \end{array} \right)\, , \  \ m \, = \, 1, \ldots, 5~, \nn\\ \Gamma_7 &=&  \left( \begin{array}{cc}  - 1_4 & 0 \\ 0 & 1_4 \end{array} \right) \, ,
\eea
where $1_4$ is the $4 \times 4$ unit matrix and $\gamma_m$ are a basis for Cliff$(5,0)$.
In this basis the dilatino condition \eqref{Dilatinodecoupled} can be written as
\bea
M \epsilon & = & 0 \, ,
\eea
where $M$ is an $8\times8$ matrix. A necessary condition
to have a non-trivial Killing spinor is $\det M=0$. We compute 
\bea
\det M &=& \frac{3^8}{2^{16}} \frac{Q^4 \bbeta^4}{ r^{14} \left(r^3+Q \right)^6} \, ,
\eea
from which we conclude that $\bbeta=0$ is necessary for supersymmetry. 
 In order to show that $\bbeta=0$ is also sufficient, we next directly 
solve the Killing spinor equation (\ref{KSEdecoupled}). 
Defining 
\begin{eqnarray}
f_1(r) & =& \frac{ r^{1/8} \sqrt{3 \sqrt{2} r^2+2 r^3+2 Q } }{\left(r^3+Q\right)^{3/8}} \, ,\nn\\
f_2(r) & =& \frac{ r^{1/8} \sqrt{-3 \sqrt{2} r^2+2 r^3+2 Q} }{\left(r^3+Q\right)^{3/8}} \, ,
\end{eqnarray}
the general solution to the dilatino and Killing spinor equation takes the form
\allowdisplaybreaks{
\begin{align}
\epsilon \ = \ \phantom{+}\sqrt{1+\sqrt{1+q^2}}
&\left(
\begin{array}{c}
 \ex^{-\frac{1}{2} \ii (\taunew+\theta +\phi +\psi )} \left(\ex^{\ii \phi } \kappa_1+\ex^{\ii \psi } \kappa_2\right) f_1(r) \\
 \ii\ex^{-\frac{1}{2} \ii (\taunew-\theta +\phi +\psi )} \left(\ex^{\ii \phi } \kappa_1-\ex^{\ii \psi } \kappa_2\right) f_1(r) \\
 \ex^{-\frac{1}{2} \ii (\taunew-\theta +\phi +\psi )} \left(\ex^{\ii (\phi +\psi )} \kappa_3+\kappa_4\right) f_2(r) \\
 \ii \ex^{-\frac{1}{2} \ii (\taunew+\theta +\phi +\psi )} \left(\ex^{\ii (\phi +\psi )} \kappa_3-\kappa_4\right) f_2(r) \\
 -\ii \ex^{-\frac{1}{2}\ii (\taunew+\theta +\phi +\psi )} \left(\ex^{\ii \phi } \kappa_1+\ex^{\ii \psi } \kappa_2\right) f_2(r) \\
 \ex^{-\frac{1}{2}\ii (\taunew-\theta +\phi +\psi )} \left(\ex^{\ii \phi } \kappa_1-\ex^{\ii \psi } \kappa_2\right) f_2(r)\\
 -\ii \ex^{-\frac{1}{2} \ii (\taunew-\theta +\phi +\psi )} \left(\ex^{\ii (\phi +\psi )} \kappa_3+\kappa_4\right) f_1(r) \\
 \ex^{-\frac{1}{2} \ii (\taunew+\theta +\phi +\psi )} \left(\ex^{\ii (\phi +\psi )} \kappa_3-\kappa_4\right) f_1(r) \\
\end{array}
\right) \nn\\
\ + \ \frac{q}{\sqrt{1+\sqrt{1+q^2}}}
&\left(
\begin{array}{c}
 \ii \ex^{-\frac{1}{2} \ii (\taunew-\theta +\phi +\psi )} \left(\ex^{\ii (\phi +\psi )} \kappa_3+\kappa_4\right) f_1(r) \\
 \ex^{-\frac{1}{2} \ii (\taunew+\theta +\phi +\psi )} \left(-\ex^{\ii (\phi +\psi )} \kappa_3+\kappa_4\right) f_1(r) \\
 -\ii \ex^{-\frac{1}{2} \ii (\taunew+\theta +\phi +\psi )} \left(\ex^{\ii \phi } \kappa_1+\ex^{\ii \psi } \kappa_2\right) f_2(r) \\
 \ex^{-\frac{1}{2} \ii (\taunew-\theta +\phi +\psi )} \left(\ex^{\ii \phi } \kappa_1-\ex^{\ii \psi } \kappa_2\right) f_2(r) \\
 \ex^{-\frac{1}{2} \ii (\taunew-\theta +\phi +\psi )} \left(\ex^{\ii (\phi +\psi )} \kappa_3+\kappa_4\right) f_2(r) \\
 \ii \ex^{-\frac{1}{2}\ii (\taunew+\theta +\phi +\psi )} \left(\ex^{\ii (\phi +\psi )} \kappa_3-\kappa_4\right) f_2(r) \\
 -\ex^{-\frac{1}{2} \ii (\taunew+\theta +\phi +\psi )} \left(\ex^{\ii \phi } \kappa_1+\ex^{\ii \psi } \kappa_2\right) f_1(r) \\
 \ii \ex^{-\frac{1}{2} \ii (\taunew-\theta +\phi +\psi )} \left(-\ex^{\ii \phi } \kappa_1+\ex^{\ii \psi } \kappa_2\right) f_1(r) \\
\end{array}
\right) \, .
\end{align}}
The four integration constants $\kappa_a$, $a=1,2,3,4$, show that the solution 
preserves half of the maximal 8 supercharges. 

When the supergravity fields are all real
the vector field
\bea\label{Killing}
K^\mu &=& \epsilon^\dagger \Gamma^\mu\epsilon
\eea
is Killing
\cite{Alday:2014bta}. In the case at hand we obtain a family of Killing vectors, depending
on the integration constants $\kappa_a$. For generic values of the parameter 
$Q$ the black hole solution has symmetry $U(1)_\taunew\times SO(4,1)$, 
where $SO(4,1)$ is the isometry group of $\mathbb{H}^4$. 
In particular this contains the maximal torus $U(1)^3\subset U(1)_\taunew\times SO(4,1)$. 
By choosing the integration constants $\kappa_a$ as
\bea
\kappa_1 &=& \frac{1}{2\sqrt{2}}\, , \qquad \kappa_2 \ = \ \kappa_3 \ = \ \kappa_4 \ = \ 0\, ,
\eea 
the Killing vector (\ref{Killing}) can be chosen to lie in the Lie algebra of this maximal torus.
Explicitly, we find
\bea\label{Ksol}
K &=& -\partial_{\phi}+\partial_{\psi}+\partial_{\taunew}\, .
\eea

\subsection{Global regularity}\label{sec:regular}

In order to have a globally regular supergravity solution we must in particular 
check that the Euclidean black hole (\ref{BH}) smoothly closes 
off at the horizon. This occurs at the largest root $r_h>0$ of the function 
$f(r)$. Imposing $f(r_h)=0$  leads to  the relation
\bea
Q &=& r_h^2\left(\frac{3}{\sqrt{2}}-r_h\right)\, .
\eea
When $Q=0$ we note that the metric (\ref{BH}) is simply  Euclidean AdS$_6$, written 
in a hyperbolic slicing, and $r_h=\frac{3}{\sqrt{2}}=\ell$ is the AdS radius. 

In general, near to $r=r_h$ the metric is to leading order
\bea
\dd s^2 & \simeq & \diff R^2 + \left(\sqrt{2}r_h-{2}\right)^2 R^2 \diff \taunew^2 + H(r_h)^{1/2}r_h^2\diff s^2_{\mathbb{H}^4}\, ,
\eea
where we have defined the new radial coordinate
\bea
R &=& 2^{3/8}3^{1/4} \frac{r_h^{1/4}}{\left(\sqrt{2}r_h-2\right)^{1/2}}(r-r_h)^{1/2}\, .
\eea
We see that the space smoothly closes off at the horizon $R=0$ 
provided $\taunew$ has period $\beta$, where
\bea
\beta &=& \frac{2\pi}{\sqrt{2}r_h-2}\, .
\eea
Comparing to section \ref{sec:fieldtheory}, where $\taunew$ has 
period $2\pi n$ with $n$ the replica index, we see that 
$\beta=2\pi n$ and
\bea\label{rh}
r_h &=& \frac{1+2n}{\sqrt{2}n}\, .
\eea
Notice that $n=1$ gives the Euclidean AdS$_6$ solution with $Q=0$.

Similarly, in order that the gauge field $\mathcal{A}$ in (\ref{XA}) is non-singular at the horizon 
we have
\bea
3\frac{H(r_h)-1}{H(r_h)}+\mu &=& 0\, ,
\eea
which using (\ref{rh}) becomes
\bea
\mu &=& -\frac{(n-1)}{n}\, .
\eea
Thus the restriction of $\mathcal{A}$ to the conformal boundary gives  
\bea
\left.\mathcal{A}\, \right|_{r=\infty} &=& \mu \dd\taunew \ = \ -\frac{(n-1)}{n}\dd\taunew \, .
\eea
Note that this agrees with the R-symmetry gauge field (\ref{AR}) required 
for supersymmetry on the $n$-branched sphere.

The resulting supergravity solution is then smooth, with the global topology 
being a product of $\R^2$ with $\mathbb{H}^4\cong \R^4$, with the origin of $\R^2$ being
the horizon at $r=r_h$. Thus the solution is defined on $\R^6$, 
with the action  of the maximal torus $U(1)^3\subset U(1)_\taunew\times SO(4,1)$ 
making this naturally into $\R^6\cong \R^2\oplus\R^2\oplus\R^2$. 
Introducing standard $2\pi$ period coordinates (\ref{Fernandoangles}),
 the Killing vector bilinear (\ref{Ksol}) becomes
\bea
K &=& b_1\partial_{\varphi_1}+b_2\partial_{\varphi_2}+b_3\partial_{\varphi_3}\, ,
\eea
where $(b_1,b_2,b_3)=\left(1,1,\tfrac{1}{n}\right)$. Also notice 
that the restriction of this vector to the conformal boundary 
at $r=\infty$ agrees with the supersymmetric Killing vector  
in section~\ref{sec:SUSY}.

\subsection{Free energy}\label{sec:freeenergy}

The holographic free energy is computed by evaluating the renormalized
on-shell action. This takes the form 
\bea
{F} &=& I_{\mathrm{ren}} \ \ = \  \ I_E + I_{\mathrm{GH}} + I_{\mathrm{counterterms}}\, .
\eea
Here $I_E$ is the Euclidean supergravity action  \eqref{Fullaction}. The Gibbons-Hawking boundary term
 is 
\bea
I_{\mathrm{GH}} &=& -\frac{1}{8\pi G_6}\int_{\partial M_6}\mathcal{K}\sqrt{\det h}\, \diff^5 x\, ,
\eea
where the space $M_6$ has boundary $\partial M_6$, $h_{mn}$ is the induced metric and 
$\mathcal{K}$ denotes the trace of the second fundamental form. 
The boundary counterterms for the general  six-dimensional Euclidean Romans $F(4)$ theory were first given in \cite{Alday:2014rxa, Alday:2014bta}. For the present case 
the two-form potential $B=0$, and consequently the general counterterm expression simplifies greatly to 
\begin{align}
I_{\mathrm{counterterms}} \ = \ \frac{1}{8\pi G_6}\int_{\partial M_6}\bigg[ &\frac{4\sqrt{2}}{3}+\frac{1}{2\sqrt{2}}R(h) +\frac{3}{4\sqrt{2}}R(h)_{mn}R(h)^{mn}-\frac{15}{64\sqrt{2}}R(h)^2 \nn \\
&-\frac{3}{4\sqrt{2}}\| \mathcal{F}\|^2_h+\frac{4\sqrt{2}}{3}(1-X)^2 \bigg]\sqrt{\det h}\,\dd^5x \, ,\label{Icounterterms}
\end{align}
where $R(h)_{mn}$, $R(h)$ are respectively the Ricci tensor and scalar of the induced metric. 
For the Euclidean black hole solution of interest the restriction of the field strength $\mathcal{F}$ to the conformal boundary is zero. 

In order to compute the regularized free energy
we cut off 
the  radial coordinate 
at $r=\Lambda$:
\bea\label{freeLambda}
{F} & = & \lim_{\Lambda \rightarrow \infty} \left[I_E(\Lambda)+ I_{\mathrm{GH}} ( \Lambda ) + I_{\mathrm{counterterms}} ( \Lambda ) \right] \, ,
\eea
where the relevant integrals are over $M_6(\Lambda)$ and $\partial M_6(\Lambda)$, respectively.
For our black hole solution 
the integrals over $\taunew$ and the hyperbolic space $\mathbb{H}^4$ 
factorize, so that the former contributes $2\pi n$ to the integral, while 
the latter contributes a factor of $\mathrm{vol}(\mathbb{H}^4)$. 
The integral over the radial variable $r$ is then easily evaluated in   
 (\ref{freeLambda}), and  we obtain
\bea
F &=& -\frac{3n}{ 4\sqrt{2} G_6}\vol(\mathbb{H}^4)\, r_h^3\, .
\eea
The volume $\vol(\mathbb{H}^4)$ is divergent. However, one can also 
regularize this using boundary counterterms (notice 
that $\mathbb{H}^4$ is Euclidean AdS$_4$). Doing so one obtains
\bea
\vol(\mathbb{H}^4) &=& \frac{4\pi^2}{3}\, .
\eea  
Substituting for the horizon radius $r_h$ in terms of $n$ (\ref{rh}), the final 
formula for the free energy is
\bea\label{Fn}
F &=& F_n \ = \ \frac{(1+2n)^3}{27n^2}F_1\, ,
\eea
where $F_1$ agrees with the free energy of Euclidean AdS$_6$ in a round 
$S^5$ slicing. This agrees precisely with (\ref{freeenergyfinal}).

\subsection{Wilson loop}

As explained in \cite{Cvetic:1999un}, solutions of the Euclidean Romans supergravity theory uplift to solutions of massive type IIA supergravity, of the warped product form $M_6\times S^4$. In \cite{Assel:2012nf} the holographic dual of a BPS Wilson loop 
in the fundamental representation was argued to be a fundamental string, 
sitting at the pole of $S^4$. Here the boundary superconformal field theories are the $USp(2N)$ gauge theories discussed in section \ref{sec:largeN}.
In \cite{Alday:2014bta} the string action for a general background was shown to be 
\bea
S_{\mathrm{string}} &=& \frac{5\pi}{4N^2G_6}\left[\int_{\Sigma_2}\left(X^{-2}\sqrt{\det \gamma}\, \diff^2 x+\ii B\right)-\frac{3}{\sqrt{2}}\mathrm{length}(\partial\Sigma_2)\right]\, .
\eea
Here $\Sigma_2$ is the string worldsheet, $\gamma_{ij}$ is the induced metric, 
and we have included a boundary counterterm to regularize the string action.

For the black hole background recall that $B=0$. We then consider 
a fundamental string $\Sigma_2\cong \R^2$ wrapping the 
$\taunew$ and $r$ directions, at a point on $\mathbb{H}^4$. 
The powers of the harmonic function $H(r)$ cancel in the integrand, 
so that
\bea
\int_{\Sigma_2}X^{-2}\sqrt{\det \gamma}\, \diff^2 x-\frac{3}{\sqrt{2}}\mathrm{length}(\partial\Sigma_2) &=&  \lim_{\Lambda\rightarrow\infty}\frac{3}{\sqrt{2}}\left(\int_{r=r_h}^{\Lambda}\diff r -\sqrt{\frac{9f(\Lambda)}{2H(\Lambda)^{3/2}}}\right)2\pi n\nn\\
&=& -\frac{6\pi n}{\sqrt{2}}r_h\, .
\eea
Identifying $-S_{\mathrm{string}}$ with $\log \, \langle W\rangle$, 
we thus find
\bea\label{Wn}
\log \, \langle W \rangle_n &=& \frac{1+2n}{3}\log \, \langle W \rangle_{n=1}\, .
\eea
Using the identification of the $n$-branched sphere partition 
function with the squashed sphere result explained in section \ref{sec:SUSY}, 
this result agrees with the large $N$ limit of the field theory computation.


\section{Discussion}\label{sec:conclusions}

In this paper we have 
computed the supersymmetric R\'enyi entropy across an entangling three-sphere
for five-dimensional superconformal field theories  using localization. 
In particular we presented a simple argument for why this 
equals the squashed five-sphere partition function, 
for appropriate squashing parameters. This argument applies 
in general dimensions.
For a class of $USp(2N)$ gauge theories we have constructed 
the holographic dual 1/2 BPS black hole solution of  Euclidean Romans $F(4)$ supergravity.
The large $N$ limit of the gauge theory result agrees perfectly with the 
supergravity computation. 

In \cite{Alday:2014rxa, Alday:2014bta} it was conjectured 
that for any supersymmetric Romans supergravity solution with
the topology of $\R^6$, with at least $U(1)^3$
isometry, and for which the Killing vector $K$ takes the
form $K=b_1\partial_{\varphi_1}+b_2\partial_{\varphi_2}+b_3\partial_{\varphi_3}$, the holographic free energy is
\bea
F &=& \frac{(|b_1|+|b_2|+|b_3|)^3}{27|b_1b_2b_3|}F_{\mathrm{AdS}_6}\, .
\eea
For the explicit 1/2 BPS black hole solution we have found in the 
present paper, the result (\ref{Fn}) agrees with this conjecture. 
Moreover, it was also conjectured that for a BPS Wilson loop 
wrapping the $\varphi_i$ circle, at the origin of the perpendicular 
$\R^4$, one has $\log \, \langle W \rangle = \frac{|b_1|+|b_2|+|b_3|}{3|b_i|} \log \, \langle 
W\rangle_{\mathrm{AdS}_6}$. Again, our result (\ref{Wn}) agrees with this formula.

Notice that the argument in section \ref{sec:SUSY} can be applied 
to any squashed sphere background. In three dimensions, 
the results of \cite{Alday:2013lba}  imply that 
the partition function for an $n$-branched \emph{squashed} three-sphere, 
with supersymmetric Killing vector  $K=b_1\partial_{\varphi_1}+b_2\partial_{\varphi_2}$,
is given by the partition function on a different squashed 
sphere with $(b_1,b_2)\rightarrow (b_1,b_2/n)$. 
Here the branch locus is the $S^1$ at $\partial_{\varphi_2}=0$. 
We expect a similar result to hold also in five dimensions. It would be
interesting  to study the implications of this for R\'enyi entropy computations.


\subsection*{Acknowledgments}

\noindent 
The work of L.~F.~A. and P.~R. is supported by ERC STG grant 306260. L.~F.~A. is a Wolfson Royal Society Research Merit Award holder.  J.~F.~S. is supported by the Royal Society. 




\begin{thebibliography}{}

\bibitem{Callan:1994py} 
  C.~G.~Callan, Jr. and F.~Wilczek,
  ``On geometric entropy,''
  Phys.\ Lett.\ B {\bf 333}, 55 (1994)
  [hep-th/9401072].
  
\bibitem{Calabrese:2004eu} 
  P.~Calabrese and J.~L.~Cardy,
  ``Entanglement entropy and quantum field theory,''
  J.\ Stat.\ Mech.\  {\bf 0406}, P06002 (2004)
  [hep-th/0405152].

\bibitem{Headrick:2010zt} 
  M.~Headrick,
  ``Entanglement Renyi entropies in holographic theories,''
  Phys.\ Rev.\ D {\bf 82}, 126010 (2010)
  [arXiv:1006.0047 [hep-th]].

\bibitem{Casini:2011kv}
  H.~Casini, M.~Huerta and R.~C.~Myers,
  ``Towards a derivation of holographic entanglement entropy,''
  JHEP {\bf 1105} (2011) 036
  [arXiv:1102.0440 [hep-th]].

\bibitem{Nishioka:2013haa}
  T.~Nishioka and I.~Yaakov,
  ``Supersymmetric R\'enyi Entropy,''
  JHEP {\bf 1310} (2013) 155
  [arXiv:1306.2958 [hep-th]].
  
\bibitem{Hung:2011nu}
  L.~Y.~Hung, R.~C.~Myers, M.~Smolkin and A.~Yale,
  ``Holographic Calculations of Renyi Entropy,''
  JHEP {\bf 1112} (2011) 047
  [arXiv:1110.1084 [hep-th]].

\bibitem{Kallen:2012va}
  J.~Kallen, J.~Qiu and M.~Zabzine,
  ``The perturbative partition function of supersymmetric 5D Yang-Mills theory with matter on the five-sphere,''
  JHEP {\bf 1208} (2012) 157
  [arXiv:1206.6008 [hep-th]].
  
\bibitem{Kim:2012ava}
  H.~C.~Kim and S.~Kim,
  ``M5-branes from gauge theories on the 5-sphere,''
  JHEP {\bf 1305} (2013) 144
  [arXiv:1206.6339 [hep-th]].

\bibitem{Hama:2011ea} 
  N.~Hama, K.~Hosomichi and S.~Lee,
  ``SUSY Gauge Theories on Squashed Three-Spheres,''
  JHEP {\bf 1105}, 014 (2011)
  [arXiv:1102.4716 [hep-th]].
  
\bibitem{Alday:2013lba}
  L.~F.~Alday, D.~Martelli, P.~Richmond and J.~Sparks,
  ``Localization on Three-Manifolds,''
  JHEP {\bf 1310} (2013) 095
  [arXiv:1307.6848 [hep-th]].

\bibitem{Alday:2014rxa}
  L.~F.~Alday, M.~Fluder, P.~Richmond and J.~Sparks,
  ``The gravity dual of supersymmetric gauge theories on a squashed five-sphere,''
  arXiv:1404.1925 [hep-th].

\bibitem{Alday:2014bta}
  L.~F.~Alday, M.~Fluder, C.~M.~Gregory, P.~Richmond and J.~Sparks,
  ``Supersymmetric gauge theories on squashed five-spheres and their gravity duals,''
  JHEP {\bf 1409} (2014) 067
  [arXiv:1405.7194 [hep-th]].

\bibitem{Imamura:2012bm} 
  Y.~Imamura,
  ``Perturbative partition function for squashed $S^5$,''
  PTEP {\bf 2013} (2013) 7,  073B01
  [arXiv:1210.6308 [hep-th]].
  
\bibitem{Ferrara:1998gv}
  S.~Ferrara, A.~Kehagias, H.~Partouche and A.~Zaffaroni,
  ``AdS(6) interpretation of 5-D superconformal field theories,''
  Phys.\ Lett.\ B {\bf 431} (1998) 57
  [hep-th/9804006].

\bibitem{Brandhuber:1999np}
  A.~Brandhuber and Y.~Oz,
  ``The D-4 - D-8 brane system and five-dimensional fixed points,''
  Phys.\ Lett.\ B {\bf 460} (1999) 307
  [hep-th/9905148].

\bibitem{Bergman:2012kr}
  O.~Bergman and D.~Rodriguez-Gomez,
  ``5d quivers and their AdS(6) duals,''
  JHEP {\bf 1207} (2012) 171
  [arXiv:1206.3503 [hep-th]].

\bibitem{Jafferis:2012iv}
  D.~L.~Jafferis and S.~S.~Pufu,
  ``Exact results for five-dimensional superconformal field theories with gravity duals,''
  JHEP {\bf 1405} (2014) 032
  [arXiv:1207.4359 [hep-th]].

\bibitem{Huang:2014gca}
  X.~Huang, S.~J.~Rey and Y.~Zhou,
  ``Three-dimensional SCFT on conic space as hologram of charged topological black hole,''
  JHEP {\bf 1403} (2014) 127
  [arXiv:1401.5421 [hep-th]].
  
\bibitem{Nishioka:2014mwa}
  T.~Nishioka,
  ``The Gravity Dual of Supersymmetric R\'enyi Entropy,''
  JHEP {\bf 1407} (2014) 061
  [arXiv:1401.6764 [hep-th]].

\bibitem{Huang:2014pda}
  X.~Huang and Y.~Zhou,
  ``N = 4 Super-Yang-Mills on Conic Space as Hologram of STU Topological Black Hole,''
  arXiv:1408.3393 [hep-th].

\bibitem{Crossley:2014oea}
  M.~Crossley, E.~Dyer and J.~Sonner,
  ``Super-R\'enyi Entropy \& Wilson Loops for N=4 SYM and their Gravity Duals,''
  arXiv:1409.0542 [hep-th].

\bibitem{Romans:1985tw}
  L.~J.~Romans,
  ``The F(4) Gauged Supergravity in Six-dimensions,''
  Nucl.\ Phys.\ B {\bf 269} (1986) 691.
  
\bibitem{Cvetic:1999un}
  M.~Cvetic, H.~Lu and C.~N.~Pope,
  ``Gauged six-dimensional supergravity from massive type IIA,''
  Phys.\ Rev.\ Lett.\  {\bf 83} (1999) 5226
  [hep-th/9906221].
  
\bibitem{Assel:2012nf} 
  B.~Assel, J.~Estes and M.~Yamazaki,
  ``Wilson Loops in 5d N=1 SCFTs and AdS/CFT,''
  Annales Henri Poincare {\bf 15}, 589 (2014)
  [arXiv:1212.1202 [hep-th]].
  
\end{thebibliography}
\end{document}